\documentclass[aps,twocolumn,showpacs,preprintnumbers,floats]{revtex4}\def\@cite#1#2{\textsuperscript{[{#1\if@tempswa , #2\fi}]}}

\usepackage{graphicx}
\usepackage{dcolumn}
\usepackage{bm}
\usepackage{epsfig}

\begin{document}

\title{$\Lambda$$\Lambda$ interactions in finite-density QCD sum rules }
\author{
 X.\ H.\ Zhong,$^{1}$\footnote {E-mail: zhongxh@ihep.ac.cn}
 P.\ Z.\ Ning$^{2}$\footnote {E-mail: ningpz@nankai.edu.cn}}
 \affiliation{$^1$Institute of High Energy Physics, Chinese Academy of Sciences, Beijing 100039, China\\
 $^2$Department of Physics, Nankai University, Tianjin 300071, China
 }

\begin{abstract}
The properties of $\Lambda$-hyperons in pure $\Lambda$ matter are
studied with the finite-density QCD sum rule approach. The first
order quark and gluon condensates in $\Lambda$ nuclear matter are
deduced from the chiral perturbation theory. The sum rule
predictions are sensitive to the four-quark condensates,
$\langle\bar{q}q\rangle^2_{\rho}$ and
$\langle\bar{q}q\rangle_{\rho}\langle\bar{s}s\rangle_{\rho}$, and
the $\pi N$ sigma term. When $\langle\bar{q}q\rangle^2_{\rho}$ is
nearly independent of density and
$\langle\bar{q}q\rangle_{\rho}\langle\bar{s}s\rangle_{\rho}$
depends strongly on density, we can obtain weakly attractive
$\Lambda$$\Lambda$ potentials (about several MeV) in low $\Lambda$
density region, which agree with the information from the latest
double $\Lambda$ hyper-nucleus experiments. The nearly no density
dependence of $\langle\bar{q}q\rangle^2_{\rho}$ and strong density
dependence of
$\langle\bar{q}q\rangle_{\rho}\langle\bar{s}s\rangle_{\rho}$ can
be explained naturally if the properties of
$\langle\bar{q}q\rangle^2_{\rho}$ and
$\langle\bar{q}q\rangle_{\rho}\langle\bar{s}s\rangle_{\rho}$ are
assumed to be similar to those of $\pi\pi$ and $\bar{K} K$ in
nuclear medium, respectively.
\end{abstract}



 \maketitle

\section{Introduction}

In neutron stars or other dense nuclear matter, $\Lambda$-hyperons
may play an important role. How to understand the interactions
between $\Lambda$-hyperons is one of interesting topics in nuclear
physics \cite{Neu,Neu1}. One can extract $\Lambda$$\Lambda$
interactions from experiments on double $\Lambda$ hyper-nuclei
\cite{Be,He,expBe,exph}. According to the latest measured data
\cite{expBe,exph}, a weak $\Lambda$$\Lambda$ potential (about
several MeV) is predicted \cite{pdepth5}, which is much weaker
than the previous predicted one, $U_{\Lambda}\simeq-20$ MeV, at
the $\Lambda$ nuclear density $0.5\rho_0$ \cite{pdepth20} with the
old measured data \cite{Be,He}. Theoretically, there have been a
few attempts towards a dynamical understanding of the
$\Lambda$$\Lambda$ interactions, such as the chiral quark model
\cite{Shen:1999pf} and the latest chiral unitary approach
\cite{LL}. However, the theoretical predictions have strong model
dependence. Further double $\Lambda$ hyper-nucleus experiments are
being carried out at KEK. There will be  more accurate
experimental data in the future, which will not only give a test
to the existing theoretical models, but also help us to develop
new methods to understand the non-perturbative $\Lambda$$\Lambda$
interactions at low energies.

In this paper, we attempt to study the $\Lambda$$\Lambda$
interactions with the finite-density QCD sum rule method (QCDSR),
which has been developed in the serial papers
\cite{sum1,sum2,sum3,sum4,sum5,sum6,sumN}. In \cite{sum5}, the
$\Lambda N$ interactions are investigated with the
$\Lambda$-hyperon in nucleonic nuclear  matter. Similarly, we
study the $\Lambda$$\Lambda $ interactions with the
$\Lambda$-hyperon in $\Lambda$ nuclear matter. We extend the
$\Lambda$ sum rules \cite{sum5} to the study of the
$\Lambda$$\Lambda $ interactions by substituting  the in-medium
condensates in nucleonic nuclear matter with those in $\Lambda$
nuclear matter.

The focus of the  finite-density QCDSR is the correlation function
of interpolating fields, which is made up of quark fields. Other
than the usual sum rules, the ground states in nuclear medium are
used for the finite-density QCDSR rather than those in the QCD
vacuum. The correlation function can be evaluated at large
space-like momenta with an operator product expansion. On the
other hand, one can obtain another presentation of the correlation
function by introducing a simple phenomenological ansatz for the
spectral densities. Finally, the sum rules can  be deduced by
equating the two different presentations of the correlation
function with appropriate weighted integrals. With the obtained
sum rules, the baryon self-energies in nuclear matter are related
to the in-medium condensates.

In this work, we only take into account the leading order
in-medium condensates. The calculations include all the
condensates up to dimension four and the first order of the
strange quark mass $m_s$. The dimension six scalar-scalar four
quark condensates are retained for they are important in the
calculations. In this work, the leading order in-medium
condensates, $\langle\bar{q}q\rangle_{\rho}$,
$\langle\bar{s}s\rangle_{\rho}$, $\langle
\frac{\alpha_s}{\pi}G^2\rangle_{\rho}$, $\langle q^{\dagger}iD_0
q\rangle_{\rho}$ and $\langle s^{\dagger}iD_0 s\rangle_{\rho}$,
are derived from the chiral perturbation theory (ChPT). We deal
with the unknown in-medium four-quark condensates $\langle
\bar{q}q\rangle^2_{\rho}$ and
$\langle\bar{q}q\rangle_{\rho}\langle \bar{s}s\rangle_{\rho}$, as
did in \cite{sum5}, by introducing two arbitrary numbers $f_1$ and
$f_2$ to describe their density dependence, respectively.

When $\langle \bar{q}q\rangle^2_{\rho}$ is nearly independent of
density (i.e., $f_1\rightarrow 0$) and
$\langle\bar{q}q\rangle_{\rho}\langle \bar{s}s\rangle_{\rho}$
depends strongly on density (i.e., $f_2\rightarrow 1$), the
$\Lambda$$\Lambda$ potential $U_{\Lambda}$ is weakly attractive,
with a value about several MeV in the low $\Lambda$ nuclear
density region (i.e., $\rho\leq0.8\rho_0$). The weakly attractive
$\Lambda$$\Lambda$ potential is compatible with the prediction
from the latest double $\Lambda$ hyper-nucleus experiments. The
weak density dependence of $\langle\bar{q}q\rangle^2_{\rho}$ and
strong density dependence of
$\langle\bar{q}q\rangle_{\rho}\langle\bar{s}s\rangle_{\rho}$ can
be explained naturally by assuming that the properties of
$\langle\bar{q}q\rangle^2_{\rho}$ and
$\langle\bar{q}q\rangle_{\rho}\langle\bar{s}s\rangle_{\rho}$ are
similar to those of $\pi\pi$ and $\bar{K} K$ in nuclear medium,
respectively.

In the subsequent section, the $\Lambda$$\Lambda$ interactions in
sum rules are given. The in-medium quark and gluon condensates are
deduced in Sec. \ref{qgc}. And the parameters are analyzed in Sec.
\ref{pan}. Then the in-medium properties versus the nuclear
density are shown in Sec. \ref{inmedium}. Finally, the summary and
conclusions are given in Sec. \ref{sc}.

\section{The $\Lambda$$\Lambda$ interactions in QCDSR}

Based on the finite-density QCDSR for the study of the $\Lambda N$
interactions in nucleonic nuclear matter, we can conveniently
extend it to the investigation of the $\Lambda$$\Lambda$
interactions in pure $\Lambda$ matter (see appendix
\ref{sumrule}). With the obtained sum rules, the scalar
self-energy $\Sigma_s$ (i.e., effective mass $M_{\Lambda}^*$ ) and
vector self-energy $\Sigma_v$ of $\Lambda$-hyperons in nuclear
medium are related to the in-medium quark and gluon condensates at
finite density. If only the in-medium quark and gluon condensates
in pure $\Lambda$ matter are determined, the self-energies
($\Sigma_s$ and $\Sigma_v$) can be obtained. Then the
$\Lambda$$\Lambda$ nuclear potential is related to the two
self-energies by the simple relation
\begin{eqnarray}
U_{\Lambda}=\Sigma_s+\Sigma_v.
\end{eqnarray}
Thus, in the subsequent section we will attempt to deduce these
quark and gluon condensates in pure $\Lambda$ matter.

\section{quark and gluon condensates} \label{qgc}
\subsection{The vacuum condensates}

Neglecting the isospin breaking effects, the vacuum condensates of
$u$, $d$ quarks are denoted by
\begin{eqnarray}
\langle \bar{u}u\rangle_{0}=\langle
\bar{d}d\rangle_{0}\equiv\langle \bar{q}q\rangle_{0}.
\end{eqnarray}
The numerical value of $\langle \bar{q}q\rangle_{0}$ can be
determined by the Gell-Mann-Oakes-Renner relation \cite{sum4}
\begin{eqnarray}
(m_u+m_d)\langle
\bar{q}q\rangle_{0}=-m_{\pi}^2f_{\pi}^2\left[1+\mathcal{O}(m_{\pi}^2)\right],
\end{eqnarray}
where $m_u$ and $m_d$ are the up and down current quark masses,
respectively; $m_{\pi}$ and $f_{\pi}$ are the pion mass and pion
decay constant, respectively. In the calculations, $m_{\pi}=138$
MeV , $f_{\pi}=93$ MeV and $m_{q}\equiv (m_u+m_d)/2=5.5 $ MeV are
adopted.

For the strange quark condensates in vacuum, we take \cite{ss,ss1}
\begin{eqnarray}
\langle \bar{s}s\rangle_{0}=0.8\langle \bar{q}q\rangle_{0},
\end{eqnarray}
the gluon condensates in vacuum are given by \cite{ss,gg}
\begin{eqnarray}
\langle \frac{\alpha_s}{\pi}G^2\rangle_{0}=(0.33 \mathrm{GeV})^4,
\end{eqnarray}
and the dimension-four quark condensates in vacuum, $\langle
q^{\dagger}iD_0q\rangle_{0}$ and $\langle
s^{\dagger}iD_0s\rangle_{0}$, are given by \cite{sum4,sum5}
\begin{eqnarray}
\langle q^{\dagger}iD_0 q\rangle_{0}=\frac{m_q}{4}\langle
\bar{q}q\rangle_{0},\\ \langle s^{\dagger}iD_0
s\rangle_{0}=\frac{m_s}{4}\langle \bar{s}s\rangle_{0},
\end{eqnarray}
where $m_s$ is the strange quark mass. In this work we adopt
$m_s=25m_{q}$ \cite{ms}.

\subsection{In-medium condensates}

The first order in-medium condensates of any
operator $\hat{\mathcal{O}}$ can be generally written as
\begin{eqnarray}\label{aa}
\langle \hat{\mathcal{O}}\rangle_{\rho}=\langle
\hat{\mathcal{O}}\rangle_{0}+\langle
\hat{\mathcal{O}}\rangle_{\Lambda}\rho+\cdot\cdot\cdot,
\end{eqnarray}
where the ellipses denote the corrections from higher orders in
density, and $\langle \hat{\mathcal{O}}\rangle_{\Lambda}$ stands
for the spin-averaged $\Lambda$ matrix element.

\subsubsection{$\langle
q^{\dagger}q\rangle_{\rho}$ and $\langle
s^{\dagger}s\rangle_{\rho}$.}

The simplest in-medium condensates are $\langle
q^{\dagger}q\rangle_{\rho}$ and $\langle
s^{\dagger}s\rangle_{\rho}$. According to the conservations of the
baryon current, one has
\begin{eqnarray}
\langle q^{\dagger}q\rangle_{\rho}=\langle
u^{\dagger}u\rangle_{\rho}=\langle
d^{\dagger}d\rangle_{\rho}=\langle
s^{\dagger}s\rangle_{\rho}=\rho.
\end{eqnarray}

\subsubsection{$\langle \bar{q}q\rangle_{\rho}$ and $\langle
\bar{s}s\rangle_{\rho}$}

The in-medium condensates $\langle \bar{q}q\rangle_{\rho}$ and
$\langle \bar{s}s\rangle_{\rho}$ can be deduced from ChPT. In the
QCD Hamiltonian density $\mathcal{H}_{\mathrm{QCD}}$, the chiral
symmetry is explicitly broken by the current quark mass terms.
This part of the Hamiltonian is given by \cite{sum2}
\begin{eqnarray}
\mathcal{H}_{mass}\equiv
m_u\bar{u}u+m_d\bar{d}d+m_s\bar{s}s+\cdot\cdot\cdot\ .
\end{eqnarray}
Neglecting the isospin breaking effects, the Hamiltonian becomes
\begin{eqnarray}
\mathcal{H}_{mass}\equiv 2m_q\bar{q}q+m_s\bar{s}s+\cdot\cdot\cdot
\ .
\end{eqnarray}
Considering the Hamiltonian $\mathcal{H}_{mass}$ as a function of
$m_q$, in the Hellmann-Feyman theorem, one obtains
\begin{eqnarray}\label{hf}
2m_q\langle \Psi(m_q) |\int dx^3  \bar{q}q|\Psi(m_q)\rangle\ \ \ \ \ \ \ \ \ \ \ \ \ \ \ \ \ \ \ \ \ \nonumber \\
=m_q\frac{d}{d m_q}\langle \Psi(m_q) | \int dx^3
\mathcal{H}_{mass}|\Psi(m_q) \rangle.
\end{eqnarray}
In Eq. (\ref{hf}), replacing $|\Psi(m_q) \rangle$ with the vacuum
state $| 0 \rangle$ and the ground state of $\Lambda$ matter $|
\rho \rangle$, respectively, then taking the difference of the two
cases, and taking into account the uniformity of the system, we
have
\begin{eqnarray}
2m_q(\langle \bar{q}q\rangle_{\rho}-\langle
\bar{q}q\rangle_{0})=m_q\frac{d\mathcal{E}}{d m_q},
\end{eqnarray}
where $\mathcal{E}$ is the energy density of $\Lambda$ matter, it
can be written as
\begin{eqnarray}
\mathcal{E}=M_{\Lambda}\rho+\delta \mathcal{E}.
\end{eqnarray}
Here $\delta \mathcal{E}$ is the contribution from the $\Lambda$
kinetic energy and $\Lambda$$\Lambda$ interactions. Because
$\delta \mathcal{E}$ is of higher order in the $\Lambda$ density
and empirically small at low densities, it is neglected in the
calculations.

Similarly, considering the Hamiltonian $\mathcal{H}_{mass}$ as a
function of $m_s$, we have
\begin{eqnarray}
m_s(\langle \bar{s}s\rangle_{\rho}-\langle
\bar{s}s\rangle_{0})=m_s\frac{d\mathcal{E}}{d m_s}.
\end{eqnarray}
In ChPT, the $\Lambda$ mass is given by (see appendix \ref{apb})
\begin{eqnarray}\label{mll}
M_{\Lambda}=M_N&+&\frac{4}{3}[(4b_D+3b_F)m_q\nonumber\\
&+&(2b_D-3b_F)m_s]B_0,
\end{eqnarray}
where $b_D$, $b_F$ and $B_0$ are real parameters in the chiral
Lagrangian, which can be well determined in the ChPT. Then we
obtain
\begin{eqnarray}
m_q\frac{dM_{\Lambda}}{dm_q}=\left [\sigma_{\pi
N}+4m_q(\frac{4}{3}b_D+b_F)B_0 \right ],
\end{eqnarray}
the $\pi N$ sigma term $\sigma_{\pi N}$ is defined as
\begin{eqnarray}
\sigma_{\pi N}\equiv m_q\frac{dM_N}{dm_q}.
\end{eqnarray}
Thus, the first-order in-medium condensate of $\langle
\bar{q}q\rangle_{\rho}$ is obtained as
\begin{eqnarray}\label{nj1}
\langle \bar{q}q\rangle_{\rho}=\langle \bar{q}q\rangle_{0}+\langle
\bar{q}q\rangle_{\Lambda}\rho,
\end{eqnarray}
where the spin averaged $\Lambda$ matrix element is
\begin{eqnarray}\label{nj2}
\langle \bar{q}q\rangle_{\Lambda}=\frac{1}{2m_q}\left [\sigma_{\pi
N}+4m_q(\frac{4}{3}b_D+b_F)B_0 \right ].
\end{eqnarray}
Comparing the condensates of $\bar{q}q$ in $\Lambda$ matter with
those in nucleonic nuclear matter, we find that there is an
additional term, $4m_q(\frac{4}{3}b_D+b_F)B_0$, in Eq.
(\ref{nj2}).

Furthermore, from the Eq. (\ref{mll}), one has
\begin{eqnarray}
m_s\frac{dM_{\Lambda}}{dm_s}=S+m_s\left[\frac{8}{3}b_D-4b_F\right]B_0,
\end{eqnarray}
where $S$ is the strangeness content in a nucleon, which is given
by \cite{sum2}
\begin{eqnarray}
 S= m_s\frac{dM_{N}}{dm_s}=\frac{y}{2}\left(\frac{m_s}{m_q}\right)\sigma_{\pi N}
 ,
\end{eqnarray}
with a dimensionless quantity
$y\equiv\langle\bar{s}s\rangle_N/\langle\bar{q}q\rangle_N$.
Analyzing the baryonic spectrum in the context of SU(3)-flavor
symmetry suggests that the strangeness content $y$ is related to
the $\pi N$ sigma term in the following manner
\cite{sigma0,sigma01}
\begin{eqnarray}
y=1-\sigma_{\pi N}^0/\sigma_{\pi N},
\end{eqnarray}
where $\sigma_{\pi N}^0$ is the $\pi N$ sigma term in the limit of
the vanishing strangeness, its value is in the range of
$\sigma_{\pi N}^0=36\pm 7$ MeV  \cite{sigma01}. The recent
determinations of the $\pi N$ sigma term suggest larger values for
it, i.e., $\sigma_{\pi N}=64\pm 8$, $(79\pm 7)$ MeV, hence $y\sim
0.5$  \cite{PN,PN1,sigma}. Thus, in this work, we adopt the new
determination of the strangeness content $y\sim 0.5$, and
constrain our discussions in the new determined region of
$\sigma_{\pi N}$, i.e., $\sigma_{\pi N}=56\sim 86$ MeV.

Finally, the first order in-medium condensates of $\langle
\bar{s}s\rangle_{\rho}$ are obtained as
\begin{eqnarray}\label{njs}
\langle \bar{s}s\rangle_{\rho}=\langle
\bar{s}s\rangle_{0}+\langle\bar{s}s\rangle_{\Lambda}\rho,
\end{eqnarray}
with the spin averaged $\Lambda$ matrix element
\begin{eqnarray}
\langle\bar{s}s\rangle_{\Lambda}=\frac{1}{m_s}\left[S+4m_s(\frac{2}{3}b_D-b_F)B_0\right].
\end{eqnarray}

\subsubsection{$\langle q^{\dagger}iD_0 q\rangle_{\rho}$ and $\langle s^{\dagger}iD_0 s\rangle_{\rho}$}

In light of Eq.~(\ref{aa}), the first order dimension-four quark
condensates in medium are written as
\begin{eqnarray}
\langle q^{\dagger}iD_0 q\rangle_{\rho}=\langle q^{\dagger}iD_0
q\rangle_{0}+\langle q^{\dagger}iD_0 q\rangle_{\Lambda}\rho,\\
\langle s^{\dagger}iD_0 s\rangle_{\rho}=\langle s^{\dagger}iD_0
s\rangle_{0}+\langle s^{\dagger}iD_0 s\rangle_{\Lambda}\rho.
\end{eqnarray}
Following Refs. \cite{sum4,sum5}, the $\Lambda$ matrix elements
$\langle q^{\dagger}iD_0 q\rangle_{\Lambda}$ and $\langle
s^{\dagger}iD_0 s\rangle_{\Lambda}$ can be related to the familiar
moments of parton distribution functions $A_{2}^{u}(\mu^2)$,
$A_{2}^{d}(\mu^2)$ and $A_{2}^s(\mu^2)$ in a $\Lambda$-hyperon as
such
\begin{eqnarray}
\langle q^{\dagger}iD_0 q\rangle_{\Lambda}&=&\frac{m_q}{4}\langle
\bar{q}q\rangle_{\Lambda}+\frac{3}{8}M_{\Lambda}A_2^{u+d}(\mu^2),\\    
\langle s^{\dagger}iD_0 s\rangle_{\Lambda}&=&\frac{m_s}{4}\langle
\bar{s}s\rangle_{\Lambda}+\frac{3}{4}M_{\Lambda}A_{2}^s(\mu^2),
\end{eqnarray}
with $A_2^{u+d}(\mu^2)=A_{2}^{u}(\mu^2)+A_{2}^{d}(\mu^2)$. The
previous studies with QCDSR predicted that
$A_{2}^{u}:A_{2}^{d}:A_{2}^{s}\simeq 0.31:0.17:0.52$ in a
$\Lambda$-hyperon at $\mu^2=1\ \ \mathrm{GeV}^2$ \cite{mom}, which
indicates that the sum of the moments for $u$, $d$ quarks are
approximately equal to those of strange quarks, i.e.,
$A_2^{u+d}\simeq A_{2}^{s}$. However, the moments do not include
the contributions of gluons, which are important in a hadron.
According to the recent predictions, the gluonic contributions to
the moments are very large, which can even reach to $\sim$0.47 in
a pion meson \cite{pig}, and $\sim$0.39 in a proton \cite{pig1}.
The moments of gluons in a $\Lambda$-hyperon should be similar to
those in the pion and proton, which are about $A^g_2\simeq 0.4$.
Momentum conservation within the $\Lambda$-hyperon is enforced by
requiring
\begin{eqnarray}
A_{2}^{u}+A_{2}^{d}+A_{2}^{s}+A^g_2=1,
\end{eqnarray}
immediately we obtain
\begin{eqnarray}
A_2^{u+d}\simeq A_{2}^{s}\simeq 0.3.
\end{eqnarray}
Although there are some uncertainties in these moments,
fortunately, they are less important to our predictions.

\subsubsection{In-medium gluon condensates}

The gluon condensates $\left\langle \frac{\alpha_s}{\pi}\left[
(u'\cdot G)^2+ (u'\cdot \tilde{G})^2\right]\right\rangle_{\rho}$
are given by \cite{sum4,sum5}
\begin{eqnarray}
\left\langle \frac{\alpha_s}{\pi}\left[ (u'\cdot G)^2+ (u'\cdot
\tilde{G})^2\right]\right\rangle_{\rho}
=-\frac{3}{2\pi}\mathcal{C}(\mu^2) M_{\Lambda}\rho,
\end{eqnarray}
where $\mathcal{C}(\mu^2)=\alpha_s(\mu^2)A_{2\Lambda}^g(\mu^2)$.
We take $\mathcal{C}(\mu^2)=0.22$ as did in \cite{sum5}
approximately. Fortunately, the predictions are nearly independent
of this kind of in-medium gluon condensate.

Another gluon condensate $\langle
\frac{\alpha_s}{\pi}G^2\rangle_{\rho}$ can be related to the trace
of the energy-momentum tensor \cite{sum2}:
\begin{eqnarray}
T^{\mu}_{\mu}=-\frac{9\alpha_s}{8\pi}G^2+2m_q\bar{q}q+m_s\bar{s}s.
\end{eqnarray}
For nuclear matter in equilibrium, the ground-state expectation
value of the trace of the energy-momentum tensor is
\begin{eqnarray} \label{t}
\langle T^{\mu}_{\mu}\rangle_{\rho}=\langle
T^{\mu}_{\mu}\rangle_{0}+\mathcal{E}.
\end{eqnarray}
Combining Eqs. (\ref{t}), (\ref{nj1}) and (\ref{njs}), we easily
obtain
\begin{eqnarray}
\langle \frac{\alpha_s}{\pi}G^2\rangle_{\rho} &=&\langle
\frac{\alpha_s}{\pi}G^2\rangle_{0}-\frac{8}{9}\{
M_{\Lambda}\nonumber\\
& & - \left [\sigma_{\pi N}+S+D_1+D_2 
\right ]\}\rho, 
\end{eqnarray}
with $D_1=4m_q(\frac{4}{3}b_D+b_F)B_0$,
$D_2=4m_s(\frac{2}{3}b_D-b_F)B_0$.

\subsubsection{$\langle
\bar{q}q\rangle^2_{\rho}$ and $\langle
\bar{q}q\rangle_{\rho}\langle \bar{s}s\rangle_{\rho}$ }

Finally, the in-medium four-quark  condensates $\langle
\bar{q}q\rangle^2_{\rho}$ and $\langle
\bar{q}q\rangle_{\rho}\langle \bar{s}s\rangle_{\rho}$ must be
considered justly in the $\Lambda$ sum rules, for they are
numerically important in the calculations. However, the in-medium
four-quark condensates in the $\Lambda$ sum rules [see Eqs.
(\ref{0}--\ref{2})] are their factorized forms, which may not be
justified in nuclear matter \cite{sum3,sum4,sumN,sum5}. Thus,
following Refs. \cite{sumN,sum5} the scalar-scalar four-quark
condensates are  parameterized so that they interpolate between
their factorized form in the QCD vacuum and in nuclear medium.
That is, in the calculations, $\langle \bar{q}q\rangle^2_{\rho}$
and $\langle \bar{q}q\rangle_{\rho}\langle \bar{s}s\rangle_{\rho}$
in Eqs. (\ref{0}--\ref{2})) are replaced with the modified forms
$\langle \widetilde{\bar{q}q}\rangle^2_{\rho}$ and $\langle
\widetilde{\bar{q}q}\rangle_{\rho}\langle
\widetilde{\bar{s}s}\rangle_{\rho}$:
\begin{eqnarray}
\langle \widetilde{\bar{q}q}\rangle^2_{\rho}&=&(1-f_{1})\langle
\bar{q}q\rangle^2_{0}+f_{1}\langle \bar{q}q\rangle^2_{\rho},\\
\langle \widetilde{\bar{q}q}\rangle_{\rho}\langle
\widetilde{\bar{s}s}\rangle_{\rho}&=&(1-f_{2})\langle
\bar{q}q\rangle_{0}\langle \bar{s}s\rangle_{0}+f_{2}\langle
\bar{q}q\rangle_{\rho}\langle \bar{s}s\rangle_{\rho},
\end{eqnarray}
where $f_{1}$ and $f_{2}$ are real parameters. Theoretically, both
$f_{1}$ and $f_{2}$ are in the range of $0\sim 1$. However, the
studies of the $\Lambda$ in nucleonic nuclear matter suggest that
$0\leq f_{1}\leq0.25$ and $0.6\leq f_{2}\leq 1$, i.e., $\langle
\bar{q}q\rangle^2_{\rho}$ depends weakly, while $\langle
\bar{q}q\rangle_{\rho}\langle \bar{s}s\rangle_{\rho}$ depends
strongly on the nuclear density \cite{sum5}.

\section{the analyses of the parameters} \label{pan}

In the calculations, we use the logarithmic measure
\cite{Measure,Measure1,sum3,sum5,sum6,sumN}
\begin{widetext}
\begin{eqnarray}
\delta(M^2)=\ln\left[\frac{\max\left\{\lambda^{*2}e^{-
(E^2_q-\textbf{q}^2)/M^2},\Pi'_{s}(M^2)/M^{*}_{\Lambda},
\Pi'_{q}(M^2),\Pi'_{u}(M^2)/\Sigma_{v}\right\}}{\min\left
\{\lambda^{*2}e^{-(E^2_q-\textbf{q}^2)/M^2},\Pi'_{s}(M^2)/M^{*}_{\Lambda},
\Pi'_{q}(M^2),\Pi'_{u}(M^2)/\Sigma_{v}\right\}}\right]
\end{eqnarray}
\end{widetext}
to quantify the fit of the left- and right- sides of the $\Lambda$
sum rules. Here $\Pi'_{s}(M^2)$, $\Pi'_{q}(M^2)$ and
$\Pi'_{u}(M^2)$ stand for the right-hand sides of the Eqs.
(\ref{0}--\ref{2}), respectively. The values of $\lambda^*$,
$s_0^*$, $M_{\Lambda}^*$ and $\Sigma_v$ are predicted by
minimizing the measure $\delta$. In the zero-density limit, we can
obtain the $\Lambda$ vacuum mass by applying the same procedure to
the sum rules.

\subsection{About the Borel mass $M^2$}

Firstly, we should choose a proper Borel mass $M^2$ in the
calculation. In principle the predictions should be independent of
the Borel mass $M^2$. However, in practice one has to truncate the
operator product expansion and use a simple phenomenological
ansatz for the spectral density, which cause the sum rules to
overlap only in some limited range of $M^2$. The previous studies
for the octet baryons show that the sum rules truncated at
dimension-six condensate do not provide a particulary convincing
plateau. Nevertheless, we can assume that the sum rules actually
has a region of overlap, although it is imperfect. Thus, in the
following we will try to find an optimization region for $M^2$, in
this region the predictions should be less sensitive to  $M^2$
than those in other regions. In Refs. \cite{sum5,sumN}, the
optimization region of $M^2$ is
suggested as 
$0.8\leq M^2\leq 1.4$ GeV$^2$, thus, in this work we choose the
proper Borel mass $M^2$ around this region.

\begin{center}
\begin{figure}[ht]
\centering \epsfxsize=8 cm \epsfbox{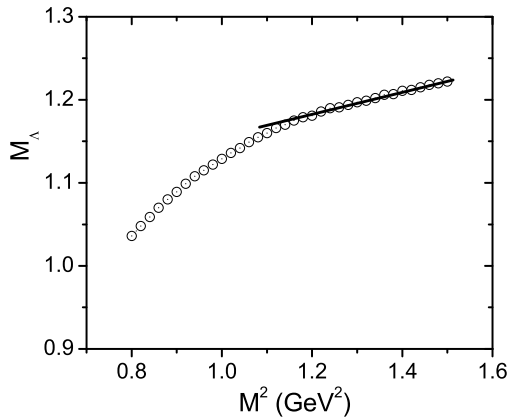}
\caption{\footnotesize $\Lambda$ mass in vacuum as a function of
$M^2$.}\label{tu1}
\end{figure}
\end{center}

To study the sensitivities of $\Lambda$ vacuum mass to $M^2$, we
plot the $\Lambda$ vacuum mass as a function of $M^2$ in the range
of $0.8\leq M^2\leq 1.5$ GeV$^2$ in Fig. \ref{tu1}. From the
figure, we see that the predicted $\Lambda$ vacuum masses are in
the range of $1036\sim 1222$ MeV, which increase monotonously with
the increment of $M^2$. In the region of $1.1\leq M^2\leq 1.5$
GeV$^2$, the predicted $\Lambda$ vacuum masses are less sensitive
to $M^2$ than those in $0.8\leq M^2\leq 1.1$ GeV$^2$, which
indicates that the optimization region of $M^2$ should be $1.1\leq
M^2\leq 1.5$ GeV$^2$.

Furthermore, we also study the sensitivities of the in-medium
properties of $\Lambda$-hyperons (e.g., $M^*_{\Lambda}$,
$\Sigma_v$ and $U_{\Lambda}$) to the Borel mass $M^2$ at the
normal nuclear density. In the calculations, we set $f_1=0.25$,
$f_2=0.8$ and $\sigma_{\pi N}=56$ MeV. To cancel the systematic
discrepancies, $M^*_{\Lambda}$, $\Sigma_v$ and $U_{\Lambda}$ are
normalized to the predicted $\Lambda$ vacuum masses.

\begin{center}
\begin{figure}[ht]
\centering \epsfxsize=8 cm \epsfbox{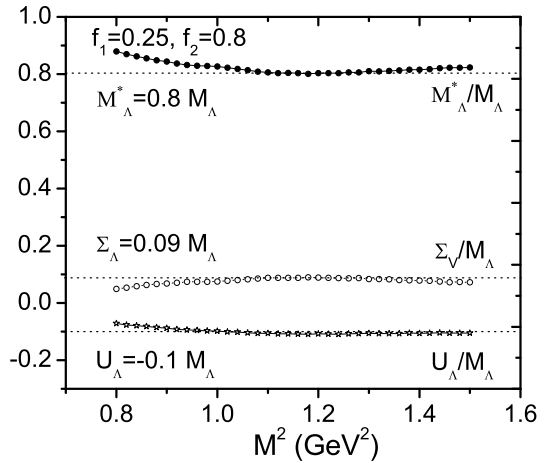} \caption{\footnotesize
The effective mass $M^*_{\Lambda}$, vector self-energy $\Sigma_v$
and the potential $U_{\Lambda}$ as functions of $M^2$ at the
density $\rho=\rho_0$.}\label{tu2}
\end{figure}
\end{center}

In Fig.\ \ref{tu2}, $M^*_{\Lambda}/M_{\Lambda}$,
$\Sigma_v/M_{\Lambda}$ and $U_{\Lambda}/M_{\Lambda}$ as functions
of $M^2$ are plotted. From the figure, we see that all the
predictions are insensitive to $M^2$ in the region of $1.1\leq
M^2\leq 1.4$ GeV$^2$, however, in the region of $0.8\leq M^2\leq
1.1$ GeV$^2$ they depend obviously on $M^2$, which also indicates
that $1.1\leq M^2\leq 1.4$ GeV$^2$ is the optimization region.

Finally, we must point out that we had better choose the lower
limit of $1.1\leq M^2\leq 1.4$ GeV$^2$ (i.e, $M^2=1.12$ GeV$^2$)
in the following calculations, for the increment of $M^2$ will
enlarge the differences between the predicted $\Lambda$ vacuum
mass and its experimental value.


\begin{center}
\begin{figure}[ht]
\centering \epsfxsize=8 cm \epsfbox{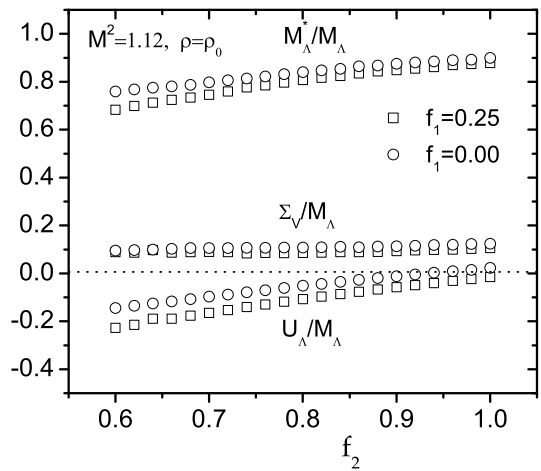} \caption{\footnotesize
The effective mass $M^*_{\Lambda}$, vector self-energy $\Sigma_v$
and the potential $U_{\Lambda}$ as functions of $f_2$ at the
density $\rho=\rho_0$.}\label{tu3}
\end{figure}
\end{center}

\subsection{The sensitivity to $f_{1}$, $f_{2}$}\label{ff}

In the calculations,  $f_{1}$ and $f_{2}$ in the parameterized
four-quark condensates are not well determined. There have been a
few discussions of them in Refs \cite{sum5,sum6}. The studies
suggest that it requires a small value of $f_{1}$ and a large
value of $f_{2}$ to obtain reasonable results. The possible
regions for $f_{1}$ and $f_{2}$ are $0\leq f_{1}\leq 0.25$ and
$0.6\leq f_{2}\leq 1.0$, respectively. In the following, the
sensitivities of the predictions to $f_{1}$ and $f_{2}$ in their
possible regions are studied.

$M^*_{\Lambda}/M_{\Lambda}$, $\Sigma_v/M_{\Lambda}$ and
$U_{\Lambda}/M_{\Lambda}$ as functions of $f_{2}$ are plotted in
Fig. \ref{tu3}. The predictions of two cases, $f_{1}=0.0$ and
$f_{1}=0.25$, are shown in the same figure, which are denoted by
circles and the squares, respectively. In the calculations,
$M^2=1.12$ GeV$^2$ and $\sigma_{\pi N}=56$ MeV are adopted.

From the figure, we find that $M^*_{\Lambda}$ and $U_{\Lambda}$
depend strongly on $f_{1}$ and $f_{2}$, however, $\Sigma_v$ is
insensitive to these parameters. The effective mass
$M^*_{\Lambda}$ increases (decreases), while the absolute value of
the potential $|U_{\Lambda}|$ decreases (increases) monotonously
with the increment of $f_{2}$ ($f_{1}$). If we fix $f_{1}=0.25$
and vary $f_{2}$ from $0.6$ to 1.0, $M^*_{\Lambda}$ increases from
$0.68 M_{\Lambda}$ to $0.88M_{\Lambda}$, and $U_{\Lambda}$
decreases from $U_{\Lambda}\simeq -0.015 M_{\Lambda}$ to
$U_{\Lambda}\simeq -0.227 M_{\Lambda}$, there is a large change
$\sim 0.2 M_{\Lambda}$ for both $M^*_{\Lambda}$ and $U_{\Lambda}$,
however, the vector self-energy has trivial changes, with a value
about $\Sigma_v=0.1 M_{\Lambda}$.

From the above analyses, we know that on condition that
$f_1\rightarrow 0$ (i.e., $\langle\bar{q}q\rangle^2_{\rho}$ is
independent of density) and $f_2\rightarrow 1$ (i.e.,
$\langle\bar{q}q\rangle_{\rho}\langle\bar{s}s\rangle_{\rho}$
depends strongly on density), we can obtain the upper limit of the
potential $U_{\Lambda}$ (i.e., the weakest potential).

\subsection{The sensitivity to $\sigma_{\pi N}$}\label{sg}

\begin{center}
\begin{figure}[ht]
\centering \epsfxsize=8 cm \epsfbox{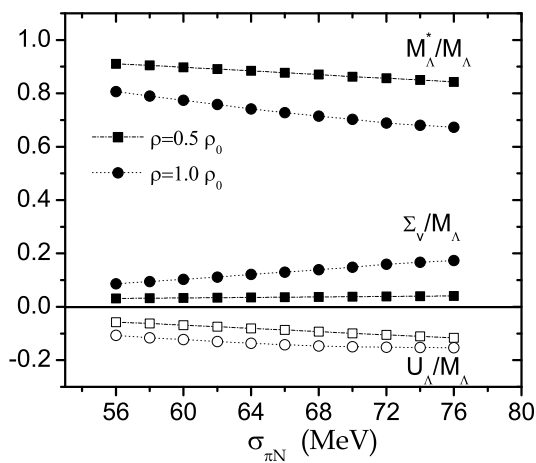}
\caption{\footnotesize The effective mass $M^*_{\Lambda}$, vector
self-energy $\Sigma_v$ and the potential $U_{\Lambda}$ as
functions of the $\pi N$ sigma term $\sigma_{\pi N}$ at the
density $\rho=\rho_0$ and $\rho=0.5\rho_0$,
respectively.}\label{tu4}
\end{figure}
\end{center}

The recent determinations of the $\pi N$ sigma term have obtained
large values: $\sigma_{\pi N}=64\pm 8$, $(79\pm 7)$ MeV. To see
the sensitivities of the predictions to $\sigma_{\pi N}$, we plot
$M^*_{\Lambda}/M_{\Lambda}$, $\Sigma_v/M_{\Lambda}$ and
$U_{\Lambda}/M_{\Lambda}$ as functions of $\sigma_{\pi N}$ in the
range of $(56\sim 76)$ MeV. In the calculations, we set $f_1=0.25$
and $f_2=0.8$. In Fig. \ref{tu4}, the predictions of two cases,
$\rho=\rho_0$ and $\rho=0.5\rho_0$, are shown (denoted by circles
and squares, respectively).

From the figure, we find that $M^*_{\Lambda}$, $\Sigma_v$ and
$U_{\Lambda}$ are more and more sensitive to $\sigma_{\pi N}$ with
the increment of the nuclear density. For example, at
$\rho=\rho_0$ obvious changes of $\Sigma_v/M_{\Lambda}$ can be
seen in the region of $\sigma_{\pi N}=(56\sim 76)$ MeV, however,
at lower density, $\rho=0.5\rho_0$, trivial changes can be seen.
The effective mass $M^*_{\Lambda}$ decreases monotonously with the
increment of $\sigma_{\pi N}$, while the vector self-energy
$\Sigma_v$ and the potential $|U_{\Lambda}|$ increase monotonously
with $\sigma_{\pi N}$. At $\rho=\rho_0$, if $\sigma_{\pi N}$
increases 2 MeV, $M^*_{\Lambda}$ will decrease a value of $\sim$20
MeV, $\Sigma_v$ and $|U_{\Lambda}|$ will increase $\sim$10 MeV,
respectively. At lower density $\rho=0.5\rho_0$, if $\sigma_{\pi
N}$ increases 2 MeV, $M^*_{\Lambda}$ and $|U_{\Lambda}|$ will
decrease $\sim$10 MeV, respectively, while $\Sigma_v$ only
increases $\sim$1 MeV.

Finally, it should be noted that when we set $\sigma_{\pi
N}\rightarrow 56$ MeV (i.e., the lower limit of the new
determinations), the upper limit of the potential $U_{\Lambda}$
(i.e., the weakest potential) is obtained.

\subsection{The sensitivity to $|\textbf{q}|$}\label{qq}

\begin{center}
\begin{figure}[ht]
\centering \epsfxsize=8 cm \epsfbox{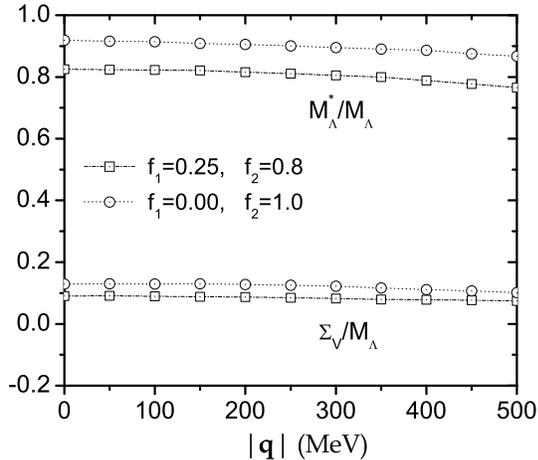} \caption{\footnotesize
The effective mass $M^*_{\Lambda}$ and vector self-energy
$\Sigma_v$ as functions of three momenta $|\textbf{q}|$ at
$\rho=\rho_0$.}\label{tu5}
\end{figure}
\end{center}

The effective mass $M^*_{\Lambda}/M_{\Lambda}$ and the vector
self-energy $\Sigma_v/M_{\Lambda}$ as functions of three momentum
$|\textbf{q}|$ at $\rho=\rho_0$ are plotted in Fig. \ref{tu5}. The
squares and circles correspond to the predictions of two cases:
$f_1=0.25$, $f_2=0.8$ and $f_1=0.0$, $f_2=1.0$, respectively.

It is shown that $M^*_{\Lambda}/M_{\Lambda}$ and
$\Sigma_v/M_{\Lambda}$ depend weakly on $|\textbf{q}|$. When
$|\textbf{q}|$ changes from zero to 500 MeV, the predicted values
of $M^*_{\Lambda}$ and $\Sigma_v$ decrease $\sim 0.06 M_{\Lambda}$
and $\sim 0.03 M_{\Lambda}$, respectively. Usually,
 $|\textbf{q}|=270$ MeV is adopted in the QCDSR calculations.

\subsection{Summary}

Now, we have known that the predictions are mainly determined by
three parameters, $f_1$, $f_2$ and $\sigma_{\pi N}$. 
The scalar self-energy $\Sigma_s$ and the potential $U_\Lambda$
are sensitive to $f_1$, $f_2$ and $\sigma_{\pi N}$. The vector
self-energy $\Sigma_v$ is insensitive to $f_1$ and $f_2$, but it
is sensitive to $\sigma_{\pi N}$ around $\rho=\rho_0$.

As a whole, at $\rho=\rho_0$, if we set $0\leq f_1\leq 0.25$,
$0.6\leq f_2\leq 1$ and $56 \leq \sigma_{\pi N}\leq 76 $ MeV, the
effective mass (scalar self-energy) and the potential $U_\Lambda$
have large possible regions, which are $M^*_{\Lambda}\simeq
(0.73\pm 0.1\pm 0.05) M_{\Lambda}$ and $U_\Lambda\simeq (-0.05 \pm
0.05 \pm 0.02) M_{\Lambda}$, respectively; the vector self-energy
$\Sigma_v$ is reasonably determined, which is $\Sigma_v\simeq
(0.14\pm 0.04) M_{\Lambda}$.

The information from the latest measured data of double $\Lambda$
hyper-nuclei indicates that the potential $U_{\Lambda}$ is weakly
attractive, with a value about several MeV. Therefore, the
potentials around the lower limit of $U_\Lambda$ ($\simeq-0.12
M_{\Lambda}$) are unreasonable. The physical predictions should be
close to the upper limit of $U_{\Lambda}$, where the parameters
$f_1\rightarrow 0$, $f_2\rightarrow 1$ and $\sigma_{\pi
N}\rightarrow 56$ MeV. The reasons why the parameters
$f_1\rightarrow 0$, $f_2\rightarrow 1$ will be discussed later.

\section{in-medium properties versus nuclear density} \label{inmedium}

\begin{center}
\begin{figure}[ht]
\centering \epsfxsize=8 cm \epsfbox{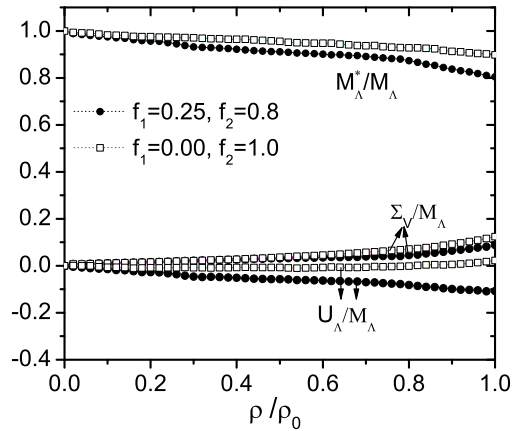}
\caption{\footnotesize The effective mass $M^*_{\Lambda}$, the
vector self-energy $\Sigma_v$ and the potential $U_{\Lambda}$ as
functions the $\Lambda$ density $\rho$.}\label{tu6}
\end{figure}
\end{center}

To further study the in-medium properties of $\Lambda$-hyperons,
we plot $M^*_{\Lambda}$, $\Sigma_v$ and $U_{\Lambda}$ as functions
of nuclear density in Fig. \ref{tu6}. On condition that $f_1=0.0$,
$f_2=1.0$ and $\sigma_{\pi N}=56$ MeV, the upper limits of
$M^*_{\Lambda}$ and $U_{\Lambda}$ at finite-density are obtained
(denoted by squares in Fig. \ref{tu6}). For comparison, the
predictions with another set of parameters, $f_1=0.25$, $f_2=0.8$
and $\sigma_{\pi N}=56$ MeV, are also presented in the same
figure, which are denoted by circles.

From Fig. \ref{tu6}, we see that the effective mass
$M^*_{\Lambda}$ decreases, while the vector self-energy $\Sigma_v$
increases  monotonously with the increment of $\Lambda$ density.
The changed tendencies agree with the predictions for baryons in
the usual Dirac Phenomenology. The two parameter sets, $f_1=0.0$,
$f_2=1.0$ and $f_1=0.25$, $f_2=0.8$, give very different
predictions for the effective mass $M^*_{\Lambda}$ and potential
$U_\Lambda$. However, $\Sigma_v$ is nearly independent of
parameter in the lower density region ($0\leq \rho\leq
0.6\rho_0$), when $\rho> 0.6\rho_0$ only weak parameter dependence
can be seen.

\begin{center}
\begin{figure}[ht]
\centering \epsfxsize=8 cm \epsfbox{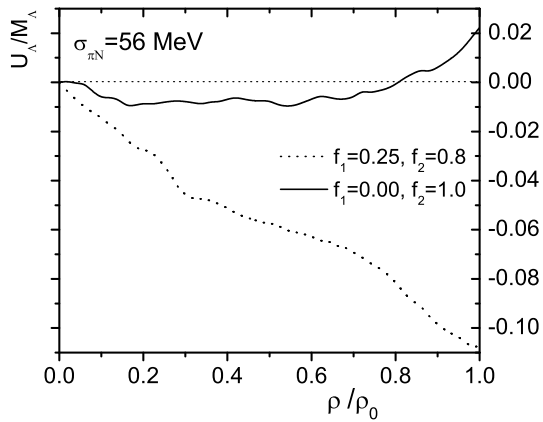}
\caption{\footnotesize The potential $U_{\Lambda}$ as a function
of the $\Lambda$ density $\rho$.}\label{tu7}
\end{figure}
\end{center}

To see the variations of potentials with $\Lambda$ nuclear density
more clearly, $U_{\Lambda}$ as a function of $\Lambda$ nuclear
density is plotted in Fig. \ref{tu7} alone. From the figure, we
find that the potentials $U_{\Lambda}$ have obvious parameter
dependence in the whole density region ($0\leq \rho\leq \rho_0$).
The differences between the two sets of parameters are more and
more obvious with the increment of $\Lambda$ nuclear density.

When we set $f_1=0.25$ and $f_2=0.8$, the potential
$|U_{\Lambda}|$ increases monotonously with the increment of
$\Lambda$ density. From Fig. \ref{tu7}, we find that with
$f_1=0.25$ and $f_2=0.8$, it gives too strong attractive
potentials $U_\Lambda$ in the whole density region, which are
inconsistent with the information from the latest double $\Lambda$
hyper-nucleus experiments.

However, with the parameters $f_1=0.0$ and $f_2=1.0$, there is a
large cancellation of the self-energies, $\Sigma_s$ and
$\Sigma_v$. In this case, we obtain the upper limit potentials,
which are weakly attractive (on the order of several MeV) in the
lower density region $\rho< 0.8 \rho_0$. There is an extremum
\begin{eqnarray}
U_{\Lambda}\simeq -0.008  M_{\Lambda}\simeq -9\ \ \mathrm{MeV}
\end{eqnarray}
around $\rho=0.5\rho_0$. Our predictions are compatible with the
latest experimental observation of the double $\Lambda$
hyper-nucleus $_{\Lambda\Lambda}^{6}$He. From the measured data,
the bound energy of $\Lambda\Lambda$ is deduced, $\Delta
B_{\Lambda\Lambda}=1.01\pm 0.20^{+0.18}_{-0.11}$ MeV \cite{expBe}.
Using the value $\Delta B_{\Lambda\Lambda}\simeq 1.01$ and
following the method of Schaffner \emph{et al.} \cite{pdepth20},
one obtains the $\Lambda$ potential $U_{\Lambda}\simeq-5$ MeV at
the density $\rho=0.5\rho_0$ \cite{pdepth5}.

From the above analyses, we know that, in order to obtain
compatible results with experiments,
$\langle\bar{q}q\rangle^2_{\rho}$ should be nearly independent of
density (i.e., $f_1\rightarrow 0$), and
$\langle\bar{q}q\rangle_{\rho}\langle\bar{s}s\rangle_{\rho}$
should depend strongly on density (i.e., $f_2\rightarrow 1.0$).

It is no accident that $\langle\bar{q}q\rangle^2_{\rho}$ depends
only weakly, while
$\langle\bar{q}q\rangle_{\rho}\langle\bar{s}s\rangle_{\rho}$
depends strongly on the nuclear density. All the finite-density
QCDSR calculations indicate that $\langle\bar{q}q\rangle^2_{\rho}$
should depend weakly on density \cite{sum3,sumN,sum5,sum6}, and
$\langle\bar{q}q\rangle_{\rho}\langle\bar{s}s\rangle_{\rho}$
should depend strongly on density \cite{sum5,sum6}. The reasons
why $\langle\bar{q}q\rangle^2_{\rho}$ has nearly no density
dependence, however,
$\langle\bar{q}q\rangle_{\rho}\langle\bar{s}s\rangle_{\rho}$ has
strong density dependence may be explained as follows:

In Ref.~\cite{Jaffe}, Jaffe have studied the possible four-quark
states. He predicted that the lowest nonet of $Q^2\bar{Q}^2$
states, C$^0(\underline{9},0^+)=u\bar{u}d\bar{d}$ and
C$^s(\underline{9},0^+)=(u\bar{u}+d\bar{d})s\bar{s}/\sqrt{2}$
(just corresponding to $\langle\bar{q}q\rangle^2_{\rho}$ and
$\langle\bar{q}q\rangle_{\rho}\langle\bar{s}s\rangle_{\rho}$),
couple strongly to $\pi\pi$ (i.e., C$^0(\underline{9},0^+)$ falls
apart into $\pi\pi$ dominatingly) and  $\bar{K}K$ (i.e.,
C$^s(\underline{9},0^+)$ falls apart into $\bar{K}K$
dominatingly), respectively. According to these predictions, we
can conclude that $\langle\bar{q}q\rangle^2_{\rho}$ and
$\langle\bar{q}q\rangle_{\rho}\langle\bar{s}s\rangle_{\rho}$
should also couple strongly to $\pi\pi$ and $\bar{K}K$ in nuclear
medium, respectively. That is, the properties of the two kinds of
in-medium four-quark condensates,
$\langle\bar{q}q\rangle^2_{\rho}$ and
$\langle\bar{q}q\rangle_{\rho}\langle\bar{s}s\rangle_{\rho}$,
should be similar to the in-medium properties of $\pi\pi$ and
$\bar{K}K$, respectively. As we know, the $\pi$ mesons, as
Goldstone bosons, do not change their properties in the nuclear
medium \cite{pion}, hence $\langle\bar{q}q\rangle^2_{\rho}$ should
be independent of the nuclear density; however, the in-medium
properties of kaon-mesons depends strongly on the nuclear density
\cite{kaon,kaon1,Kaplan,brown,c1,c4,c5}, hence
$\langle\bar{q}q\rangle_{\rho}\langle\bar{s}s\rangle_{\rho}$ has
strong density dependence.

In the calculations, the minimum measure $\delta$ is at the order
of $10^{-6}\sim 10^{-5}$. The parameters $\lambda^*$ and $s_0^*$
have density dependence. For example, with $f_1=0.25,\ \ f_2=0.8$,
when the density increases from zero to $\rho_0$ the optimized
value for the residue $\lambda^*$ will decrease from
$3.27\times10^{-2}$ GeV$^3$ to $1.66\times10^{-2}$ GeV$^3$, and
the optimized value for the continue threshold $s_0^*$ will
decrease  from 2.86 GeV$^2$ to 1.94 GeV$^2$.

\section{summary and conclusions}\label{sc}

Based on the finite-density QCDSR for describing the $\Lambda N$
interaction in nucleonic nuclear matter, we conveniently extend it
to the study of the $\Lambda$$\Lambda$ interactions in $\Lambda$
nuclear matter. The in-medium condensates of
$\langle\bar{q}q\rangle_{\rho}$, $\langle\bar{s}s\rangle_{\rho}$,
$\langle \frac{\alpha_s}{\pi}G^2\rangle_{\rho}$, $\langle
q^{\dagger}iD_0 q\rangle_{\rho}$ and $\langle s^{\dagger}iD_0
s\rangle_{\rho}$ are derived from the ChPT.

The $\Lambda$ potentials $U_\Lambda$ are sensitive to the
in-medium four quark condensates and the $\pi N$ sigma term (i.e.,
three parameters $f_1$, $f_2$ and $\sigma_{\pi N}$), for the
scalar self-energies $\Sigma_s$ are sensitive to them. There is a
large cancellation of the scalar self-energy $\Sigma_s$ and vector
self-energy $\Sigma_v$, each is on the order of a few hundred MeV
around $\rho=\rho_0$. On condition that $f_1\rightarrow 0$ (i.e.,
$\langle\bar{q}q\rangle^2_{\rho}$ is independent of density),
$f_2\rightarrow 1$ (i.e.,
$\langle\bar{q}q\rangle_{\rho}\langle\bar{s}s\rangle_{\rho}$
depends strongly on density) and $\sigma_{\pi N}\rightarrow 56$
MeV (i.e., the lower limit of the new determinations), the upper
limit of the $\Lambda$ nuclear potentials are predicted, which are
weakly attractive (about several MeV) in low density region $\rho<
0.8 \rho_0$. In this case, the predicted $\Lambda$ nuclear
potentials agree well with the latest experimental observation of
a double $\Lambda$ hyper-nucleus $_{\Lambda\Lambda}^{6}$He.

The nearly density independent $\langle\bar{q}q\rangle^2_{\rho}$
and strongly density dependent
$\langle\bar{q}q\rangle_{\rho}\langle\bar{s}s\rangle_{\rho}$ can
be explained naturally by assuming the properties of
$\langle\bar{q}q\rangle^2_{\rho}$ and
$\langle\bar{q}q\rangle_{\rho}\langle\bar{s}s\rangle_{\rho}$ are
similar to those of $\pi\pi$ and $\bar{K} K$ in nuclear medium,
respectively.

In this work, the $\pi N$ sigma term in the new determinations
(i.e., $\sigma_{\pi N}=56$ MeV) are adopted, and hence a large
strange content of the nucleon (i.e., $y=0.5$) are obtained
according to the overviews in \cite{PN}. The reasonable results
predicted by us support these new determinations.

It is a preliminary attempt to study the $\Lambda\Lambda$
interactions in finite $\Lambda$ density with QCDSR. More studies
are needed to extend QCDSR to finite density. The four-quark
condensates in medium should be studied further. In light of our
predictions, how to relate the four-quark condensates to two
pseudoscalar mesons in the practical calculations should be
considered carefully in our later work.

\section*{  Acknowledgements }
We would like to thank Prof. Q. Zhao and G.X.Peng for helpful
discussions and good suggestions. This work is supported, in part,
by China Postdoctoral Science Foundation, the institute of high
energy physics, CAS, and the Natural Science Foundation of China
(Grand No. 10575054).

\appendix

\section{the $\Lambda$ sum rules} \label{sumrule}
The sum rules for the $\Lambda$ hyperon  propagating in the
nuclear matter had been deduced by Xuemin Jin and R. J.
Furnstahl\cite{sum5}, which are given by
\begin{widetext}
\begin{eqnarray}
\lambda^{*2}M_{\Lambda}^*e^{-(E^2_q-\textbf{q}^2)/M^2}
&=&-\frac{m_s}{48\pi^4}M^6E_2L^{-8/9}-\frac{M^4}{12\pi^2}E_1\left(4\langle
\bar{q}q\rangle_{\rho}-\langle
\bar{s}s\rangle_{\rho}\right)-\frac{m_s}{6\pi^2}\bar{E}_qM^2E_0(\langle
q^{\dagger}q\rangle_{\rho}-\langle
s^{\dagger}s\rangle_{\rho})L^{-8/9}\nonumber\\
&&-\frac{2m_s}{9\pi^2}\textbf{q}^2(2\langle
q^{\dagger}iD_0q\rangle_{\rho}+\langle s^{\dagger}iD_0
s\rangle_{\rho})L^{-8/9}+\frac{4m_s}{3}\langle
\bar{q}q\rangle_{\rho}^2\nonumber\\
&&-\frac{8}{27}\langle \bar{q}q\rangle_{\rho}(m_s\langle
\bar{s}s\rangle_{\rho} +2\langle s^{\dagger}iD_0
s\rangle_{\rho})-\frac{16}{27}\left(1-\frac{\textbf{q}^2}{M^2}\right)\langle
\bar{q}q\rangle_{\rho}(m_s\langle \bar{s}s\rangle_{\rho}-4\langle
s^{\dagger}iD_0 s\rangle_{\rho})\nonumber\\
&&-\frac{8}{9}\bar{E}_q\langle \bar{q}q\rangle_{\rho}\langle
s^{\dagger} s\rangle_{\rho}+\frac{4}{9}\bar{E}_q\langle
q^{\dagger} q\rangle_{\rho}\langle\bar{s}s\rangle_{\rho},\label{0}\\
\lambda^{*2}e^{-(E^2_q-\textbf{q}^2)/M^2}
&=&\frac{M^6}{32\pi^4}E_2L^{-4/9}+\frac{M^2}{32\pi^2}\langle
\frac{\alpha_s}{\pi}G^2\rangle_{\rho}E_0L^{-4/9}-\frac{m_s}{3\pi^2}M^2E_0\langle
\bar{q}q\rangle_{\rho}L^{-4/9}\nonumber\\
&&+\frac{M^2}{144\pi^2}\left(E_0-4\frac{\textbf{q}^2}{M^2}\right)\left\langle
\frac{\alpha_s}{\pi}\left[ (u'\cdot G)^2+ (u'\cdot
\tilde{G})^2\right]\right\rangle_{\rho}L^{-4/9}\nonumber\\
&&+\frac{m_s}{54\pi^2}M^2\left(17E_0-2\frac{\textbf{q}^2}{M^2}\right)\langle\bar{s}s\rangle_{\rho}L^{-4/9}-
\frac{4M^2}{27\pi^2}\left(2E_0-5\frac{\textbf{q}^2}{M^2}\right)\langle
q^{\dagger}iD_0q\rangle_{\rho} L^{-4/9}\nonumber\\
&&-\frac{M^2}{27\pi^2}\left(7E_0-4\frac{\textbf{q}^2}{M^2}\right)\langle
s^{\dagger}iD_0s\rangle_{\rho}L^{-4/9}+\frac{\bar{E}_q}{18\pi^2}M^2E_0(5\langle
q^{\dagger}q\rangle_{\rho}+\langle
s^{\dagger}s\rangle_{\rho})L^{-4/9}\nonumber\\
&&-\frac{2}{9}(\langle \bar{q}q\rangle_{\rho}^2-4\langle
\bar{q}q\rangle_{\rho}\langle
\bar{s}s\rangle_{\rho})L^{-4/9}+\frac{4}{9}(2\langle
q^{\dagger}q\rangle_{\rho}^2+\langle
q^{\dagger}q\rangle_{\rho}\langle s^{\dagger}s\rangle_{\rho})L^{-4/9},\label{1}\\
\lambda^{*2}\Sigma_{v}e^{-(E^2_q-\textbf{q}^2)/M^2}
&=&\frac{1}{36\pi^2}M^4E_1(11\langle
q^{\dagger}q\rangle_{\rho}+13\langle
s^{\dagger}s\rangle_{\rho})L^{-4/9}\nonumber\\
&&-\frac{\bar{E}_q}{27\pi^2}M^2E_0(7m_s\langle\bar{s}s\rangle_{\rho}
-32\langle q^{\dagger}iD_0q\rangle_{\rho}-28\langle
s^{\dagger}iD_0s\rangle_{\rho})L^{-4/9}\nonumber\\
&&-\frac{\bar{E}_q}{36\pi^2}M^2E_0\left\langle
\frac{\alpha_s}{\pi}\left[ (u'\cdot G)^2+ (u'\cdot
\tilde{G})^2\right]\right\rangle_{\rho}L^{-4/9}+\frac{4\bar{E}_q}{9}\langle
q^{\dagger}q\rangle_{\rho}^2L^{-4/9}\nonumber\\
&&+\frac{20\bar{E}_q}{9}\langle q^{\dagger}q\rangle_{\rho}\langle
s^{\dagger}s\rangle_{\rho}L^{-4/9}\label{2}.
\end{eqnarray}
\end{widetext}
Naturally we can conveniently extend these sum rules to the study
of the $\Lambda$-hyperon in pure $\Lambda$ matter by changing the
in-medium condensates in nucleon matter to the in-medium
condensates in $\Lambda$ matter. In Eqs. (\ref{0}--\ref{2}),
$\langle \hat{\mathcal{O}} \rangle_{\rho}$ stands for the
condensate of a general operator $\hat{\mathcal{O}}$ in the
$\Lambda$ matter, such as $\langle \bar{q}q\rangle_{\rho}$ stands
for the dimension-three quark condensate, $\langle
q^{\dagger}iD_0q\rangle_{\rho}$ stands for the dimension-four
quark condensate,  $\langle \frac{\alpha_s}{\pi}G^2\rangle_{\rho}$
and $\left\langle \frac{\alpha_s}{\pi}\left[ (u'\cdot G)^2+
(u'\cdot \tilde{G})^2\right]\right\rangle_{\rho}$ stand for two
gluon condensates. $\lambda^*$ is the residue at the
quasi-lambda-hyperon pole, and $M$ is known as the Borel mass. The
quantities account for continuum corrections to the sum rules are
defined as:
\begin{eqnarray}
E_0 &\equiv&
 \left(1-e^{-s_0^*/M^2}\right), \\
E_1 &\equiv&
 \left[1-e^{-s_0^*/M^2}\left( \frac{s_0^*}{M^2}+1\right)\right],\\
E_2 &\equiv&
 \left[1-e^{-s_0^*/M^2}\left( \frac{s_0^{*2}}{2M^4}+\frac{s_0^*}{M^2}+1\right)
 \right],
\end{eqnarray}
with the continue threshold
\begin{eqnarray}
s_0^*=\omega_0^2-\textbf{q}^2,
\end{eqnarray}
where, $\omega_0$ is the energy at the continuum threshold, and
$|\textbf{q}|$ is the three-momentum of the quasi-Lambda-hyperon.
In Eqs. (\ref{0}--\ref{2}) we have defined
\begin{eqnarray}
M_{\Lambda}^*&\equiv& M_{\Lambda}+\Sigma_s, \\
E_{q}&\equiv&\Sigma_v+\sqrt{M_{\Lambda}^{*2}+\textbf{q}^2},\\
\bar{E}_{q}&\equiv& \Sigma_v-\sqrt{M_{\Lambda}^{*2}+\textbf{q}^2},
\end{eqnarray}
where, $\Sigma_s$ and $\Sigma_v$ are the scalar and vector
self-energies of the $\Lambda$ hyperon in $\Lambda$ matter,
respectively. $M^*$ is the $\Lambda$ effective mass in $\Lambda$
matter. $E_{q}$ and $\bar{E}_{q}$ correspond to the positive- and
negative-energy poles, respectively.

The factor $L$ in Eqs. (\ref{0}--\ref{2}) is defined as
\begin{eqnarray}
L\equiv \frac{\ln M/\Lambda_{QCD}}{\ln \mu/\Lambda_{QCD}},
\end{eqnarray}
where $\mu$ is the normalization point of the operator product
expansion, and  $\Lambda_{QCD}$ is the QCD scale parameter. In
numerical calculations, one takes $\Lambda_{QCD}=0.1\
\mathrm{GeV},\ \ \mu=0.5\ \mathrm{GeV}$ \cite{mu}.

\section{the hadron mass in ChPT}\label{apb}

In chiral perturbation theory, the hadron masses originate in the
chiral breaking. The leading term of the explicitly chiral
breaking Lagrangian for mesons is
\begin{eqnarray}\label{L1}
\mathcal{L}^{\phi}_{sb}=B_0\frac{f^2}{2}\langle
\mathcal{M}(U+U^{\dagger})\rangle,
\end{eqnarray}
where $B_0=-\langle \bar{q}q\rangle_{0}/f^2_{\pi}$ is the order
parameter of spontaneous symmetry violation. $\mathcal{M}$ is the
quark mass matrix $\mathcal{M}$=diag$(m_q,m_q,m_s)$. The
pseudoscalar meson decay constants are equal in the $SU(3)_V$
limit and denoted by $f = f_ {\pi}$. $U=\exp(i\sqrt{2}\phi/2)$, in
which $\phi$ collects the pseudoscalar meson octet. The explicitly
chiral breaking Lagrangian for baryons is given by \cite{Chpt}
\begin{eqnarray}\label{L2}
\mathcal{L}^{B}_{sb}&=& M_0\langle \bar{B}B\rangle+4B_0b_0\langle
\bar{B}B\rangle\langle \mathcal{M}\rangle\nonumber\\
&+&4B_0b_D\langle\bar{B}\{\mathcal{M},B\}\rangle+4B_0b_F\langle\bar{B}[\mathcal{M},B]\rangle,
\end{eqnarray}
where $M_0$ is the common octet baryon mass in the chiral limit,
$B$ is the ground state SU(3) baryon octet consisting of the
nucleons and hyperons which are collected in a $3\times3$ matrix,
and $b_0$, $b_D$ and $b_F$ are the parameters to be determined.
From the chiral Lagrangian, we can get the masses for $\pi$, $K$
and $\eta$
\begin{eqnarray}
m_{\pi}^2&=&2m_qB_0,\label{mp}\\
m_K^2&=&(m_q+m_s)B_0,\label{mk}\\
m_{\eta}^2&=&\frac{2}{3}(m_q+2m_s)B_0,
\end{eqnarray}
and the masses of different baryons \cite{Chpt},
\begin{eqnarray}
M_{\Lambda}&=&\tilde{M}_{0}-\frac{4}{3}(m_K^2-m_{\pi}^2)b_D,\label{mn}\\
M_N&=&\tilde{M}_{0}-4m_K^2b_D+4(m_K^2-m_{\pi}^2)b_F,\label{ml}\\
M_{\Sigma}&=&\tilde{M}_{0}-4m_{\pi}^2b_D,\label{ms}\\
M_{\Xi}&=&\tilde{M}_{0}-4m_{K}^2b_D-4(m_{K}^2-m_{\pi}^2)b_F.
\end{eqnarray}
From the relations, (\ref{mn}-\ref{ms}), we can obtain the
parameters $b_D$ and $b_F$, which are determined by
\begin{eqnarray}
b_D&=&\frac{3}{4}(M_{\Sigma}-M_{\Lambda})/(m_K^2-4m_{\pi}^2),\\
b_F&=&b_D +\frac{1}{4}(M_{N}-M_{\Sigma})/(m_K^2-m_{\pi}^2).
\end{eqnarray}
Combining Eqs. (\ref{mp}, \ref{mk}) and (\ref{mn}, \ref{ml}), one
has
\begin{eqnarray}\label{mlll}
M_{\Lambda}=M_N&+&\frac{4}{3}[(4b_D+3b_F)m_q\nonumber\\
&+&(2b_D-3b_F)m_s]B_0.
\end{eqnarray}


\begin{thebibliography}{99}

\bibitem{Neu}
J. Schaffner and I. N. Mishustin, Phys. Rev. C \textbf{53}, 1416
(1996).
\bibitem{Neu1}
T. Takatsuka, S. Nishizaki, Y. Yamamoto, R. Tamagaki, Prog. Theor.
Phys. \textbf{115}, 355 (2006).
\bibitem{Be} M. Danysz, et al., Phys. Rev. Lett. \textbf{11}, 29 (1963);
 M. Danysz et al., Nucl. Phys. \textbf{49}, 121 (1963).
\bibitem{He} D. J. Prowse, Phys. Rev. Lett. \textbf{17}, 782 (1966).
\bibitem{expBe}H. Takahashi et al., Phys. Rev. Lett. \textbf{87}, 212502 (2001).
\bibitem{exph} J. K. Ahn et al., Phys. Rev. Lett. \textbf{87}, 132504 (2001).

\bibitem{pdepth5} H. Q. Song , R. K. Su , D. H. Lu and W. L. Qian,  Phys. Rev. C \textbf{68}, 055201 (2003).
\bibitem{pdepth20} J.Schaffner ,C. B. Dover, A. Gal  \emph{et al.},  Ann. Phys. \textbf{235}, 35 (1994).

\bibitem{Shen:1999pf}
  P.~N.~Shen, Z.~Y.~Zhang, Y.~W.~Yu, X.~Q.~Yuan and S.~Yang,
  J.\ Phys.\ G {\bf 25}, 1807 (1999).
\bibitem{LL} K. Sasaki, E. Oset, and M. J. Vicente Vacas,
Phys.Rev.C\textbf{74}, 064002 (2006).
\bibitem{sum1} T. D. Cohen, R. J. Furnstahl, and D. K. Griegel,
Phys. Rev. Lett. \textbf{67}, 961 (1991).
\bibitem{sum2} T. D. Cohen, R. J. Furnstahl, and D. K. Griegel,
Phys. Rev. C\textbf{45}, 1881 (1992).
\bibitem{sum3} R. J. Furnstahl, D. K. Griegel, and T. D. Cohen,
Phys. Rev. C\textbf{46}, 1507 (1992).
\bibitem{sum4} X. Jin, T. D. Cohen, R. J. Furnstahl, and D. K. Griegel,
Phys. Rev. C\textbf{47}, 2882 (1993).
\bibitem{sumN} X. Jin, M. Nielsen, T. D. Cohen, R. J. Furnstahl, and D. K. Griegel,
Phys. Rev. C\textbf{49}, 464 (1994).
\bibitem{sum5} X. Jin, R. J. Furnstahl, Phys. Rev. C\textbf{49}, 1190 (1994).
\bibitem{sum6} X. Jin, M. Nielsen, Phys. Rev. C\textbf{51}, 347 (1995).






\bibitem{ss} M. A. Shifman, A. I. Va\"{\i}nshte\"{\i}n and V. I. Zakharov, Nucl. Phys.
\textbf{B147}, 448 (1979); \textbf{B147}, 519 (1979);

\bibitem{ss1} B. L. Ioffe, Nucl. Phys. \textbf{B188}, 317 (1981) [Erratum-ibid.
 \textbf{B191}, 591 (1981) ]; L. J. Reinders, H. Rubinstein and S.
Yazaki, Phys. Rep. \textbf{127}, 1 (1985).

\bibitem{gg} M. A. Shifman, A. I. Va\"{\i}nshte\"{\i}n and V. I. Zakharov, Nucl. Phys.
\textbf{B147}, 385 (1979).

\bibitem{ms} J. Gasser and H. Leutwyler, Phys. Rep. \textbf{87}, 77 (1982); S.
Weinberg, Trans. New York Acad. Sci. \textbf{38}, 185 (1977); H.
Leutwyler, Phys. Lett. \textbf{B378}, 313 (1996).
%
\bibitem{sigma0} J. Gasser, H. Leutwyler and M. E. Sainio, Phys. Lett. B \textbf{253}, 252 (1991).
%
\bibitem{sigma01} Marc Knecht ,PiN Newslett. \textbf{15}, 108 (1999).
%
\bibitem{PN} J. Ellis, Eur.Phys.J. A\textbf{24}, s2, 3 (2005)[
hep-ph/0411369].
%
\bibitem{PN1} M. M. Pavan, I. I. Strakovsky, R. L. Workman and
R. A. Arndt, PiN Newslett. \textbf{16}, 110 (2002); T. Inoue, V.
E. Lyubovitskij, T. Gutsche and A. Faessler, Phys. Rev.
C\textbf{69}, 035207 (2004) and references therein.
%
\bibitem{sigma} P. Schweitzer, Eur. Phys. J. A \textbf{22}, 89 (2004); R. Koch, Z.
Phys. C \textbf{15}, 161 (1982); J. Gasser, H. Leutwyler, M.E.
Sainio, Phys. Lett. B \textbf{253}, 252; 260 (1991).
%
\bibitem{mom} V. L. Chernyak, A. A. Ogloblin and I. R. Zhitnitsky,
Z.Phys.C\textbf{42}, 569, (1989).
%
\bibitem{pig} S. Falciano et al. (NA10), Z. Phys. C\textbf{31}, 513
(1986);  K. Wijesooriya, P.E. Reimer, R.J. Holt,  Phys. Rev.
C\textbf{72}, 065203 (2005).
\bibitem{pig1} A. D. Martin, R. G. Roberts, W. J. Stirling, and R. S.
Thorne, Eur.Phys.J. C\textbf{4}, 463 (1998).






\bibitem{Measure} V. M. Belyaev and B. L. Ioffe, Zh. Eksp. Teor.
Fiz. \textbf{83}, 876 (1982)[Sov. Phys. JETP \textbf{56}, 493
(1982)]; \textbf{84}, 1236 (1983) [\textbf{57}, 716 (1983)].
\bibitem{Measure1} D. B. Leinweber, Ann. Phys.\textbf{198}, 203 (1990).


\bibitem{Jaffe} R. J. Jaffe, Phys. Rev. D\textbf{15}, 267 (1977).


\bibitem{pion} W. Cassing and E. L. Bratkovskaya, Phys. Rep. \textbf{308}, 65 (1999).

\bibitem{kaon} X. H. Zhong, G. X. Peng, and P. Z. Ning, Phys. Rev. C\textbf{72}, 065212 (2005).

\bibitem{kaon1} X. H. Zhong, G. X. Peng, L. Li, and P. Z. Ning, Phys. Rev. C\textbf{74}, 034321 (2006).

\bibitem{Kaplan}
D. B. Kaplan and A. E. Nelson, Phys. Lett. B \textbf{175}, 57
(1986).

\bibitem{brown}
G. E. Brown and C. -H. Lee \emph{et al.},
 Nucl. Phys. A\textbf{567}, 937 (1994).

\bibitem{c5}
G. Q. Li, C. -H. Lee, and G.E. Brown,
 Nucl. Phys. A\textbf{625}, 372 (1997).

\bibitem{c4}
J. Schaffner, I. N. Mishustin, and J. Bondorf,
 Nucl. Phys. A \textbf{625}, 325 (1997).


\bibitem{c1}
J. Schaffner, A. Gal, I.N. Mishustin, H. St\H{o}cker, and W.
Greiner,
 Phys. Lett. B \textbf{334}, 268 (1994).

\bibitem{mu}B. L. Ioffe and A. V. Smilga, Nucl. Phys. \textbf{B232}, 109
(1984).

\bibitem{Chpt} B. Borasoy, Nucl. Phys. A\textbf{754}, 191c (2005).

\end{thebibliography}
\end{document}